\documentclass{ws-procs975x65}
\usepackage{multirow}

\begin{document}

\title{Properties of the propagating oscillatory shock wave in the accretion flows around few 
transient black hole candidates during their outbursts}

\author{DIPAK DEBNATH$^{1*}$, SANDIP. K. CHAKRABARTI$^{2,1}$}

\address{1. Indian Centre For Space Physics, 43 Chalantika, Garia Station Road, Kolkata, 700084, India\\
2. S.N. Bose National Center for Basic Sciences, JD-Block, Salt Lake, Kolkata, 700098, India\\
}

%

\begin{abstract}

In our study of the timing properties of few Galactic black hole candidates evolutions of the low and intermediate 
frequency quasi-periodic oscillations (LIFQPOs) are observed. 
In 2005, for explaining evolution of QPO frequency during rising phase of 2005 GRO~J1655-40 outburst, Chakrabarti 
and his students introduced a new model, namely propagating oscillatory shock (POS) model. Here we present 
the results obtained from the same POS model fitted QPO evolutions during both the rising and declining phases 
of the outbursts of 2005 GRO~J165540, 2010-11 GX~339-4, and 2010 \& 2011 H~1743-322.

\end{abstract}

\keywords{Black Holes, shock waves, accretion disks, Stars:individual (GRO~J1655-40, GX~339-4, H~1743-322)}

\bodymatter

\section{Introduction}

Generally, most of the transient black hole candidates (BHCs) show 
LIFQPOs ranging between $0.01$ to $30$~Hz in their power density spectra. 
More precisely these QPOs are observed during hard and intermediate (hard-intermediate or soft-intermediate) 
spectral states. Our study during the outburst phases of few Galactic transient black hole candidates (for 
e.g., GX~339-4, H~1743-322, GRO~J1655-40, XTE~J1550-564 etc.) show that during the rising phases QPO 
frequencies are observed to be increasing monotonically and during declining phases the sources show 
a monotonically decreasing nature of QPO frequencies. More precisely from the spectral study, it was 
noticed that these evolutions of QPO frequency are observed during hard and hard-intermediate spectral 
states \cite{DD10,Nandi12,DD13}.


\section{Observation and Data Analysis}

We used RXTE/PCA data of GRO~J1655-40, GX~339-4 \& H~1743-322 outbursts with a maximum timing 
resolution of $8 ms - 125\mu s$ to generate power-density spectra (PDS), using ``powspec" task 
of XRONOS package of HeaSOFT 6.11 with a normalization factor of `-2' to have the `white' noise 
subtracted {\it rms} fractional variability in $2-15$ keV (0-35 channels of PCU2) light curves 
of $0.01$~sec time bins. To find centroid frequency of QPOs, PDS are fitted with Lorentzian profiles. 

\section{Results}

\subsection{Origin of QPO and SOM model}

Observations of LIFQPOs in black hole candidates are reported quite extensively in the literature and 
at the same time many theoretical or empirical models 
are introduced to explain this important temporal feature of BHCs.
But one satisfactory model namely shock oscillation model (SOM) by Chakrabarti and his collaborators \cite{MSC96} 
in mid 90s, shows that the oscillation of X-ray intensity is actually due to the oscillation of the post-shock (Comptonizing) 
region. 
In this SOM solution, at the rising phase of the outburst, a shock wave moves toward the black hole, 
and at the declining phase of the outburst, a shock moves away from the black hole, which oscillates 
either because of a resonance (where the cooling time of the flow is approximately the infall time; 
\cite{MSC96}) or because the Rankine-Hugoniot condition is not satisfied \cite{RCM97} to form a steady shock.

\begin {figure}[t]
\vskip -0.5 cm
\centering{
\includegraphics[scale=0.6,angle=0,width=4.0truecm]{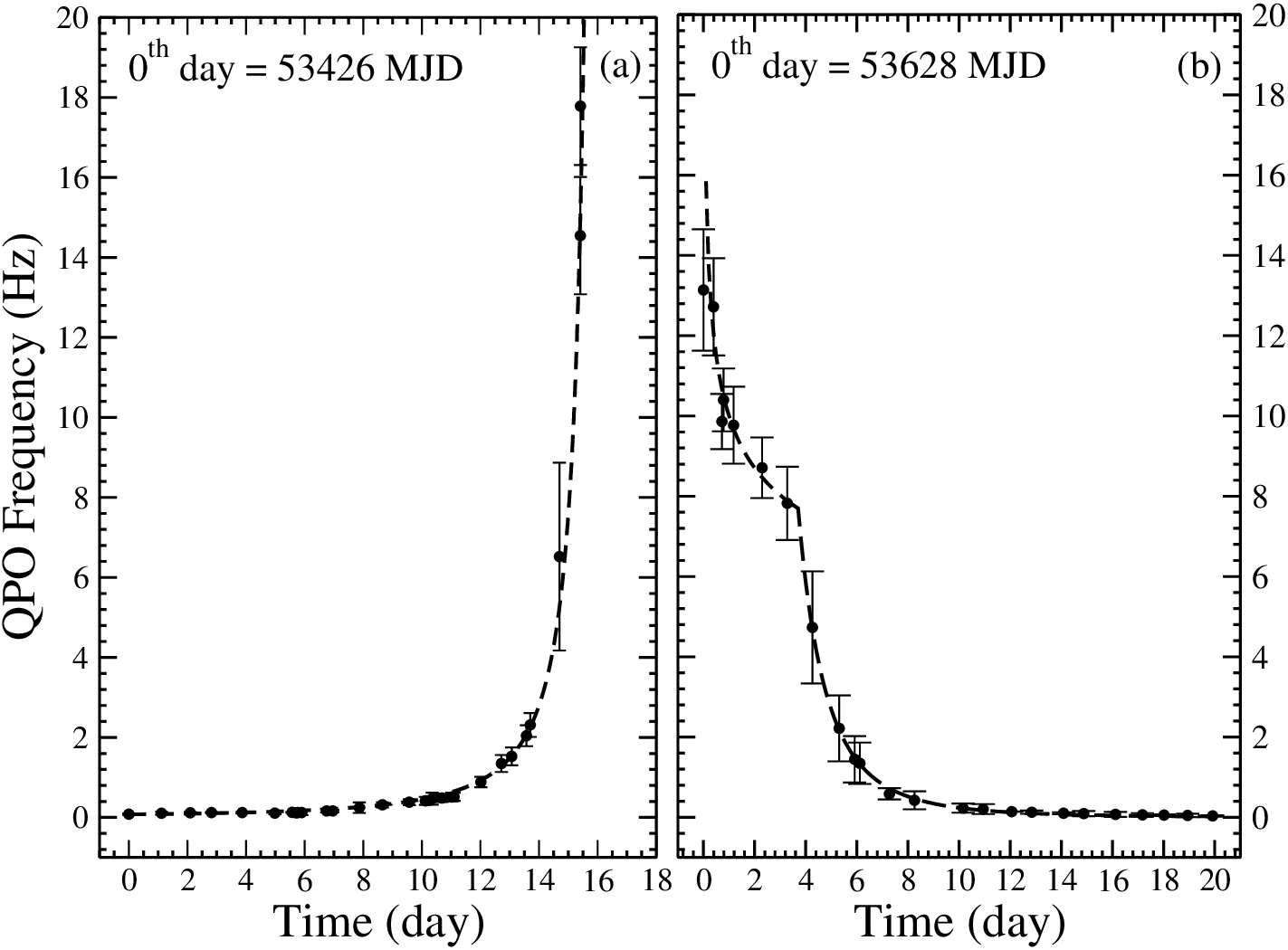}\hskip 0.15cm
\includegraphics[scale=0.6,angle=0,width=4.0truecm]{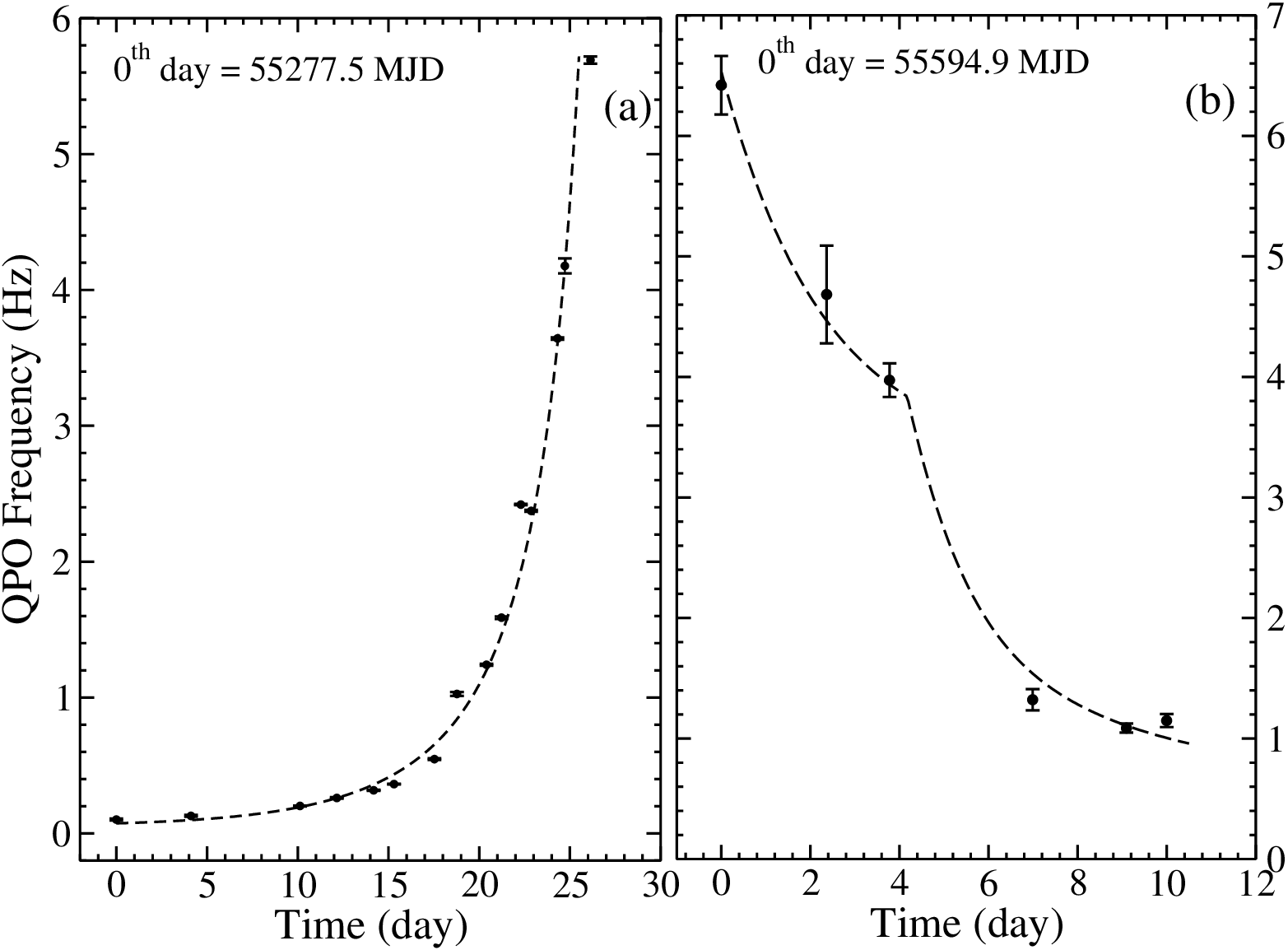}\hskip 0.25cm
\includegraphics[scale=0.6,angle=0,width=4.0truecm]{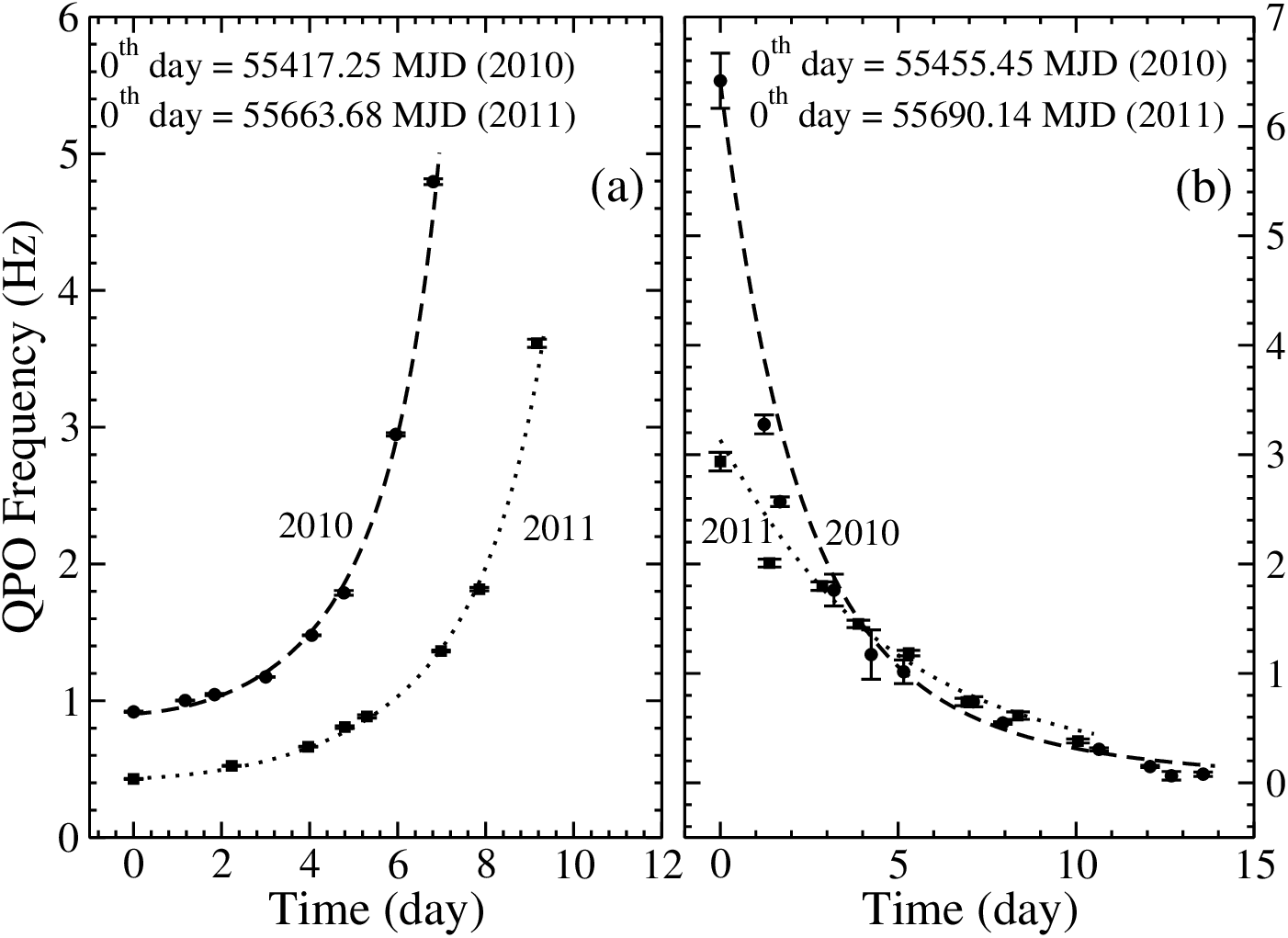}
}
\caption{Variations of the QPO frequency with time (in day) of (a) the rising phases and (b) the declining 
phases of 2005 outburst of GRO~J1655-40\cite{C08} (in left panel), 2010-11 outburst of GX~339-4\cite{DD10,Nandi12} 
(in middle panel), and combined 2010 \& 2011 outbursts of H~1743-322\cite{DD13} (in right panel) are fitted 
with POS model (dashed/dotted curves).}
\label{kn : fig1}
\end {figure}

\subsection{Evolution of QPO frequency and POS model}

For explaining evolution of the QPO frequency during the rising phase of 2005 GRO~J1655-40 outburst, 
Chakrabarti and his students \cite{C05} introduced a new model, namely propagating oscillatory shock (POS) 
model. According to POS, the QPO frequency is inversely proportional to the infall time ($t_{infall}$) 
in post-shock region. For governing equations of POS model, see Eq.1-3 of Chakrabarti et al. (2008)\cite{C08}.

So far, we have studied QPO frequency evolutions during the outbursts of GRO~J1655-40 (2005)\cite{C05,C08}, 
GX~339-4 (2010-11)\cite{DD10, Nandi12}, H~1743-322 (2010 \& 2011)\cite{DD13}, and XTE~J1550-564 (1998)\cite{C09}. 
These evolutions are well fitted with POS model and from there accretion flow parameters related to evolutions 
are extracted. Here, we discuss a brief summary of model fitted results obtained from the QPO frequency evolutions 
of first three sources during their individual X-ray outbursts.

\subsubsection{2005 outburst of GRO~J1655-40}

During the rising phase, 
QPO frequency was increased monotonically from $82$mHz to $17.78$Hz. According to the POS model fit, 
shock moved from $1200~r_g$ to $59~r_g$ with a constant velocity ($v$) of $\sim 20~m~s^{-1}$ with $R_0=4$ and $\alpha=0.003$.
During the declining phase, 
QPO frequency was decreased monotonically from $13.14$Hz to $34$mHz and this evolution was fitted 
with the same POS model. Here, we observed that initially (upto $3.5$ days) shock moved very slowly 
(from $40~r_g$ to $59~r_g$) and then rapidly with high acceleration (on last QPO observed day at $3100~r_g$). 
During the entire period, strength of the shock was kept constant with $R=4$.

\subsubsection{2010-11 outburst of GX~339-4}

Evolution of QPO frequency from $102$mHz to $5.69$Hz observed during the rising phase of the outburst, 
where shock was observed to move from $1500~r_g$ to $172~r_g$ with a constant velocity ($v$) of 
$\sim 10~m~s^{-1}$, $R_0=4$, and $\alpha=0.00068$. 
During the declining phase, QPO frequency is found to decrease from $6.42$Hz to $1.149$Hz. 
According to POS, initial $4.2$ days shock moved away slowly (from $84$ to $155~r_g$) with initial 
velocity ($v_i$) = $205~cm~sec~^{-1}$ and acceleration =$20~cm~sec^{-1}~day^{-1}$. After that as of 
GRO~J1655-40, shock disappears with higher rate of acceleration ($175~cm~sec^{-1}~day^{-1}$) with final 
velocity ($v_f$) = $1785~cm~sec^{-1}$.

\subsubsection{2010 and 2011 outbursts of H~1743-322}

During rising phase of 2010 outburst, 
QPO frequency was increased from $0.92$ to $4.79$Hz, with the movement of shock wave 
from $428$ to $181~r_g$ with $R_0$=1.39, $\alpha=0.006$, and an accelerating velocity, 
changing from $180$ to $1133$ cm/sec within a period of $6.81$ days. Similarly, during 2011 
rising phase, QPO frequency was increased from $0.43$ to $3.61$Hz with movement 
of shock wave from $550$ to $217~r_g$ with the help of an accelerating velocity  
(changing $340$ - $1137$ cm/sec) within a period of $9.16$ days, $R_0=2$, $\alpha=0.0055$.
During declining phase of 2010 outburst, QPOs were observed to decrease from $6.42$Hz to $79$mHz 
with an accelerating velocity changing from $560$ to $1578$ cm/sec and movement of shock wave from $65$ 
to $751~r_g$ within a period of $13.57$ days, and $R=2$ (constant). Similarly, during the 2011 declining 
phase, shock was found to move from $118$ to $411~r_g$ with an accelerating velocity (changing $460$ - $912$ cm/sec) 
and constant $R=2.78$. As a result QPO frequency was observed to decrease from 
$2.94$ to $0.38$Hz with in a period of $10$ days.

\section{Discussions and Concluding Remarks}

From the successful interpretation of evolutions of QPO frequency during rising and declining phases of 
transient BHCs with the POS solution \cite{C08,C09,DD10,Nandi12,DD13}, we are certain that one can 
calculate the frequency of QPOs if one knows the instantaneous shock locations or vise-versa.
We hope that this model will also be able to explain the evolutions of QPO frequency for other transient BHCs.

\end{document}